%% file: eeIe_v0.tex
\documentclass{iopart}

\usepackage{graphicx, dblfloatfix}
\usepackage{iopams}
\usepackage{hyperref}
\usepackage{cite}
\usepackage{color, xcolor}
\usepackage{rotating}
\usepackage{booktabs}
\usepackage{multirow}
\usepackage{amssymb}
\usepackage{array}

\expandafter\let\csname equation*\endcsname\relax
\expandafter\let\csname endequation*\endcsname\relax
\usepackage{amsmath}

\include{my_macro}

\newcommand{\phib}{\phi_\mrm{b}}
\newcommand{\phip}{\phi_\mrm{p}}
\newcommand{\Ap}{A_\mrm{w}}

\newcommand{\Ie}{I_\mrm{e}}
\newcommand{\Iesat}{I_\mrm{es}}

\newcommand{\CS}{C_\mrm{S}}
\newcommand{\Teff}{T_\mrm{eff}}
\newcommand{\mi}{m_\mrm{i}}

\newcommand{\np}{n_\mrm{p}}

\newcommand{\Te}{T_\mrm{e}}
\newcommand{\Ti}{T_\mrm{i}}
\newcommand{\me}{m_\mrm{e}}

\newcommand{\DebL}{\lambda_\mrm{De}}

\newcommand{\vthe}{v_\mrm{th,e}}
\newcommand{\Iestrunc}{I_\mrm{es,T}}
\newcommand{\Iesmax}{I_\mrm{es,M}}
\newcommand{\Iestran}{I_\mrm{es,tr}}

\newcommand{\Iesinf}{\Iesat^\mrm{inf}}
\newcommand{\IesEP}{\Iesat^\mrm{EP}}

\newcommand{\Iesinfcurv}{\left. \Iesinf \right|_\mrm{curv}}
\newcommand{\IesEPcurv}{\left. \IesEP \right|_\mrm{curv}}

\newcommand{\IesEPext}{\left. \IesEP \right|_\mrm{ext}}
\newcommand{\kape}{\kappa_\mrm{e}}
\newcommand{\kapetrunc}{\kappa_\mrm{eT}}
\newcommand{\kapemax}{\kappa_\mrm{eM}}
\newcommand{\phipinf}{\phip^\mrm{inf}}
\newcommand{\phipmxc}{\phip^\mrm{mxc}}
\newcommand{\phipmxf}{\phip^\mrm{mxf}}
\newcommand{\phipep}{\phip^\mrm{EP}}
\newcommand{\phiphyp}{\phip^\mrm{hyp}}
\newcommand{\nuIE}{\nu_{\mrm{IE}}^{e{\text -}e}}
\newcommand{\nuLB}{\nu_{\mrm{LB}}^{e{\text -}e}}

\begin{document}
\title[]{Will shifted and truncated Maxwellian EVDF near the wall exhibit different electron currents?}

\date{\today}

\author{Yegeon~Lim$^{1,2\dagger}$, and Y.-c.~Ghim$^{2\ast}$\footnote{Author to whom any correspondence should be addressed.}}
\address{$^1$Applied Physics and Materials Science, California Institute of Technology, Pasadena, CA 91125, United States of America}
\address{$^2$Department of Nuclear and Quantum Engineering, Korea Advanced Institute of Science and Technology, Daejeon 34141, Republic of Korea}
\ead{\mailto{$^\dagger$yglim@caltech.edu}, \mailto{$^\ast$ycghim@kaist.ac.kr}}


\begin{abstract}

Experiments to explore the predicted consequences on the \textit{I--V} characteristics of a Langmuir probe based on the fluid approach for electrons and the electron Bohm criterion are conducted and their results are presented. The predictions on the \textit{I--V} characteristics of a Langmuir probe when electrons are highly collisional are shown in terms of the estimation of plasma potentials and electron saturation currents, primarily represented by a rounded knee of the curve. An edge-effect reduced and guard-ringed Langmuir probe is employed to minimize geometrical effects on the measured data. To characterize the rounded knee of the curve, various methods on defining critical points around the plasma potential are presented. Additionally, emissive and cutoff probes were utilized to obtain bulk plasma properties, providing accurate reference plasma parameters. The physically valid condition for electrons to exhibit a shifted Maxwellian velocity distribution near the wall, wherein the fluid approach is applicable, is established from theory of the ion-acoustic instability enhanced electron-electron collisions. In this work, we conducted experiments scanning from collisionless to collisional regimes for electrons with theoretical considerations. Our observation indicate that the Langmuir probe can be understood with the conventional wisdom represented by the model of truncated electron velocity distribution. This result suggests the absence of the enhancement of electron-electron collisions and/or incompleteness of electron fluid approach near the wall.

\end{abstract}

\maketitle 
\ioptwocol

\section{Introduction}\label{sec:ee_intro}
The electron current towards the wall from the bulk plasmas is well described by the kinetic model associated with a non-Maxwellian electron velocity distribution function (EVDF)\cite{Langmuir_1925_RN984}. This kinetic model, often referred to as the `truncated EVDF model', assumes collisionless electrons resulting in the finite electron flux toward the wall by which the electrons having sufficient kinetic energy to overcome electric potential in the space charge region formed in front of the negatively biased wall from the plasma potential. As the electrons over the threshold velocity are absorbed by the wall, their velocity distribution is considered as truncated function. If the wall bias is equal to the plasma potential, the electron flux collected by the wall is a random flux. The flux of the electrons collected by the wall can be expressed by the first moment of the truncated EVDF. In the case of a Maxwellian distributed EVDF at the bulk, and for $\phib\le\phip$ where $\phib$ is the wall bias and $\phip$ is the plasmas potential, the electron flux $\Gamma_\mrm{e}$ is \cite{Hershkowitz:1989}
\begin{equation}
	\Gamma_\mrm{e}=\kape n_\mrm{e}\vthe\exp\left[-\frac{(\phip-\phib)}{\Te}\right],
	\label{eq:ee_Ie}
\end{equation}
where $n_\mrm{e}$ denotes electron density, $\vthe$ thermal velocity of the electrons $\vthe\equiv\sqrt{e\Te/\me}$, $e$ the electron charge, $\Te$ electron temperature in eV, and $m_\mrm{e}$ the electron mass. The constant factor $\kape$ has a value of $\kapetrunc\equiv(2\pi)^{-0.5}\approx0.4$ where the subscript T refer to the model with `truncation'. Equation (\ref{eq:ee_Ie}) has been widely accepted as a conventional wisdom and validated through numerous experimental studies.

As electrons are not always be considered collisionless, the kinetic model with truncated EVDF may not properly describe the electron collection. The other extreme case with collisional electrons must have shifted Maxwellian velocity distribution in accordance with H-theorem. By making an additional assumption of isothermal electron temperature, we can approach with a two-fluid picture to properly express the electron current. Guittienne \textit{et al.}\cite{Guittienne_2018_RN137} have arrived at the same form of expression for $\Ie$ as in equation (\ref{eq:ee_Ie}), except the constant $\kapetrunc$ replaced by $\kapemax\approx0.6$ where the subscript M stands for shifted `Maxwellian'. In the two-fluid approach, $\kapemax$ represents the density factor at the sheath edge compared to the bulk, rather than the resultant factor from the truncated Maxwellian, $\kapetrunc$. This is in line with previous studies on electron sheath ($\phib\geq\phip$) theory to the extend that we consider the case $\phib\approx\phip$, which establish the electron Bohm criterion\cite{Loizu_2012_RN101, Scheiner_2015_RN491, Scheiner_2016_RN536, Yee_2017_RN158, Sun_2022_RN986}, i.e., $u_\mrm{e,se}\ge\vthe$, where $u_\mrm{e,se}$ represents the electron fluid speed towards the wall at the electron sheath edge.

In typical low temperature plasmas having electron temperature $\Ti\ll\Te\sim1$~eV and plasma density $\np\lesssim10^{9}$~cm$^{-3}$, the electron-electron collision mean free path $\lambda_\mrm{e{\text -}e}^\mrm{LB}$, calculated via the classical Lenard-Balescu collision operator\cite{Nicholson}, is $\sim10$~m which is typically much larger than the dimension of vacuum chambers. On the other hand, the electron-neutral mean free path $\lambda_\mrm{e{\text -}n}$ has, for instance, the order of magnitude of $\sim1$~m for Ar discharges at $1$~mTorr, and is inversely proportional to the neutral density ($\propto n_\mrm{g}^{-1}$). Thus, the collisionality of electrons is commonly considered to be governed by collisions between electrons and neutral particles in low temperature plasmas. However, the corresponding length scale of the presheath is approximately $3$~cm, and is scaled by ion-neutral mean free path $\lambda_\mrm{i{\text -}n}$ which is also inversely proportional to the neutral density. In this example, the given corresponding value of electron mean free path is much larger than the presheath length scale, and their comparable order of magnitudes holds for most of the different discharge conditions. Therefore, with the classical picture, electrons in low temperature plasmas are commonly considered to be collisionless with regard to physical phenomena on plasma--wall interactions as the presheath is considered largest spatial region bridging the bulk and the wall.

Instabilities in plasmas can be source of mechanism on increasing effective collisionality of charged particles. Baalrud \textit{et al.} claimed through his theoretical works\cite{Baalrud_2009_RN346, Baalrud_2009_RN715} that the collisions between charged particles are enhanced by ion-acoustic instability. The ion-acoustic instability is established by finite relative flows between ions and neutral particles. Situations of such condition are common in the ion sheath and presheath region for low temperature plasmas. The presence of the instability enhanced ion--ion collisions are experimentally identified in the works associated with the generalized Bohm criterion for multiple ion species plasmas \cite{Yip_2010_RN923, Baalrud_2015_RN637, Yip_2015_RN164, Severn_2017_RN63}. The theory argue that the electron--electron collisions can also be enhanced by the instability. Therefore, by properly setting up the experimental conditions, the physical situation with the Maxwellian EVDF shifted towards the wall may be achieved facilitated by the ion-acoustic instability.

In two- or three-dimensional situation, the EVDF may be considered as a shifted Maxwellian attributed by a geometric or pressure gradient effects. Scheiner \textit{et al.}\cite{Scheiner_2015_RN491} showed through his 2D particle-in-cell simulations that the shifted Maxwellian EVDF has been observed near the biased wall for collisionless electrons which is originated by the finite size of electron collecting wall. The presence of the wall causes electron shadowing invoking a loss-cone type of EVDF leading to the shifted Maxwellian-like distribution when integrated with respect to velocity components parallel to the wall surface. Furthermore, the flow of the electrons towards the sheath for $\phib\gtrsim(\phip-\Te)$ is found to driven by a pressure gradient that slightly shifts the maximum peak of the EVDF \cite{Scheiner_2015_RN491, Scheiner_2016_RN536}.

Recently, Scheiner developed a theory\cite{Scheiner_2024_RN1095} with the effusion based losses of the electrons to the wall that can describe his observation on the EVDF and the formation of the electron presheath. The theory results in that the electron flux with the shifted Maxwellian-like EVDF originated by electron shadowing or pressure gradient effects is consistent with the conventional description, the random flux of electrons to the wall. The author also noted that the electron sheath Bohm criterion\cite{Scheiner_2015_RN491, Scheiner_2016_RN536, Yee_2017_RN158} only applies to situation of very collisional electrons, i.e., a true shifted Maxwellian EVDF.

Jin \textit{et al.}\cite{Jin_2022_RN913} conducted experiments to confirm the electron sheath theory that will give a different consequence in the electron saturation current compared to the description with random flux of electrons. They compared ion and electron currents using an asymmetric double probe method in a condition with collisionless electrons. Consistent with Scheiner's theory\cite{Scheiner_2024_RN1095}, the naturally formed shifted Maxwellian-like EVDF due to the geometric effect of the probe do not affect the conventional wisdom that electron sheath collects a random flux of electrons when $\phib\approx\phip$.

In this work, we seek for experimental evidences that can support the predictions in accordance with theories that suppose a shifted Maxwellian EVDF. We use a custom-designed edge-effect reduced Langmuir probe (EERP) to obtain data for singly ionized Ar plasmas in a multidipole chamber with a filament discharge source\cite{Lim_2020_RN239, Lim_2024_RN961}. The experiments were conducted under conditions where the ion-acoustic instability can be established near the EERP to sufficiently enhance electron--electron collisions to the extent that the EVDF becomes shifted Maxwellian. We focus on the properties observed around the knee of the \textit{I--V} characteristics measured by the EERP. To obtain the bulk plasma parameters, we utilize supplementary diagnostics such as an emissive probe and a cutoff probe to acquire reference values for plasma potential and density, respectively, taking advantage of the uniform plasmas in the multidipole chamber.

Section \ref{sec:ee_expect} presents our predictions on \textit{I--V} characteristics for the experiments described in section \ref{sec:ee_setup}. Examples of the EERP data and methods on estimating critical points around the plasma potential are provided in section \ref{sec:ee_Vp}. The experimental results and discussions on them are presented in section \ref{sec:ee_results}, and a conclusion of this work in section \ref{sec:ee_conclusion}.

\section{Expectations on \textit{I-V} characteristics in the presence of electron--electron collisions}\label{sec:ee_expect}
We will employ the theoretically predicted ion-acoustic instability-enhanced (IE) electron--electron collisional process\cite{Baalrud_2009_RN715}. The collision frequency is proportional to the classical Coulomb collision frequency $\nuLB$ which increases exponentially toward the wall from the onset of instability ($x=0$):
\begin{equation}
	\nuIE \sim \frac{\nuLB}{8 \ln \Lambda} \frac{1+2 \kappa_{c}^{2}}{\left(1+\kappa_{c}^{2}\right)^{2}} \exp \left(\eta \frac{x}{l}\right),
	\label{eq:IE}
\end{equation}
where
\begin{align*}
&\nuLB\sim\frac{\omega_\mrm{pe}}{8\pi\np\DebL^3} \ln\Lambda,\\
&\eta\equiv\frac{l}{\DebL}\sqrt{\frac{\pi m_e}{8m_i}},\\
&\kappa_c\equiv\left\{
\begin{array}{cl}
	\sqrt{\CS^2/u_\mrm{i}^2-1}\hfill\text{  for }u_\trm{i}<\CS,\\
	0\hfill\text{  for }u_\trm{i}\ge\CS.
\end{array}\right.
\end{align*}
Here, $\Lambda$ is the Coulomb logarithm ($\approx10$), $\omega_\mrm{pe}$ the electron plasma frequency, $\DebL$ the Debye length, $\me$ and $\mi$ are the the electron and ion masses, respectively, $\CS\equiv\sqrt{\Te/\mi}$ the ion sound speed, $u_\mrm{i}$ the ion fluid speed. The value $l$ is a length scale characterizing the presheath, typically the mean free path of ion--neutral collisions. As the solution of the presheath is valid for $\sim2l$ from the wall\cite{Oksuz_2002_RN125}, it is considered that the instability grows over the length $\sim2l$, thus the collision frequency at the sheath edge can be estimated by setting $x=2l$, i.e., $\nuIE(x=2l)$.

\begin{figure}\centering
\includegraphics[width=\linewidth]{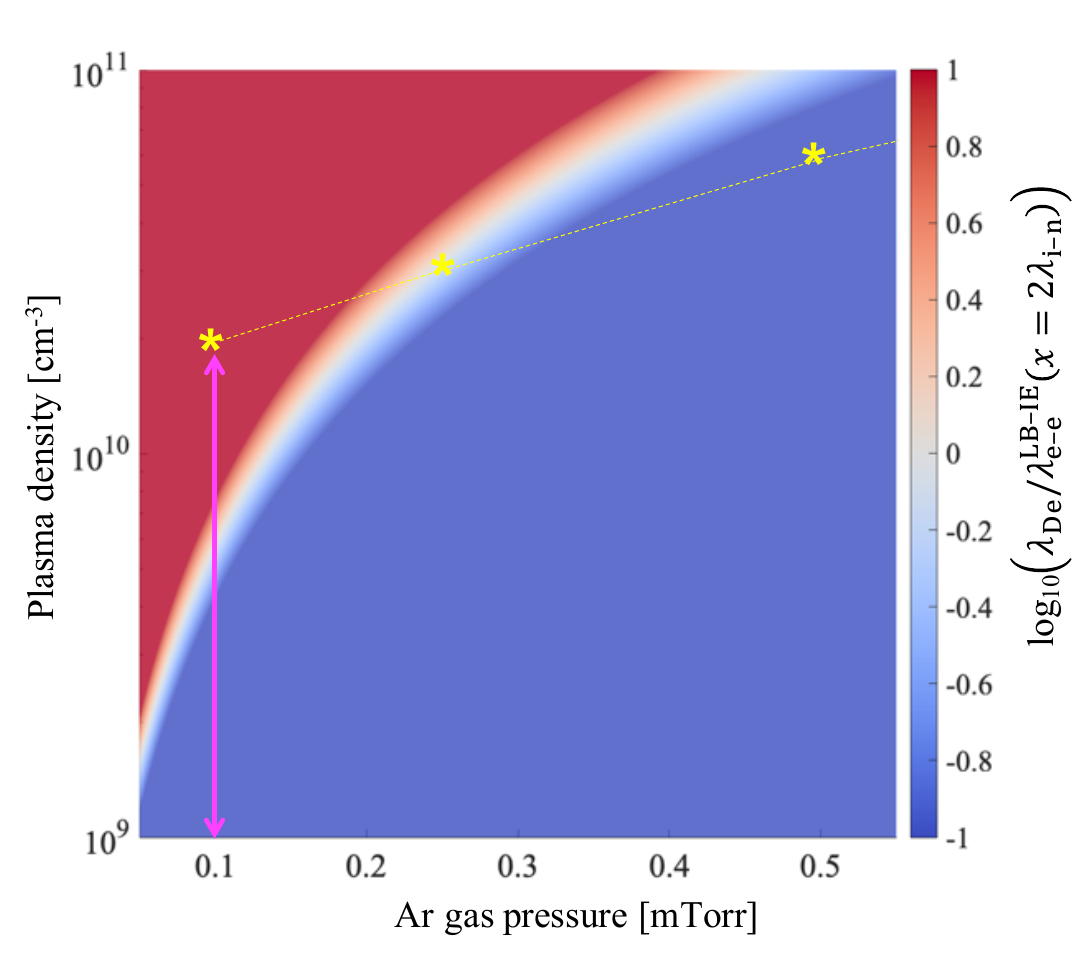}
\caption{The calculated ratio of the Debye length $\DebL$ to e--e mean free path $\lambda_\mrm{e{\text -}e}^\mrm{LB+IE}$ as a function of plasma density and Ar neutral pressure for Ar plasmas, with $\Te=1$~eV is assumed. The yellow asterisks represent the upper limit of the plasma density that can be achieved in our multidipole plasma device for the corresponding operation pressure.}
\label{fig:IE}
\end{figure}

Figure \ref{fig:IE} illustrates the calculated ratio of the Debye length to the mean free path of electron--electron collisions $\lambda_\mrm{e{\text -}e}^\mrm{LB+IE}=\vthe/(\nuIE+\nuLB)$ expected at the sheath edge for Ar plasmas, considering both classical and IE collisions with an electron temperature of $\Te=1$~eV. Note that electron--neutral collisions are negligible for our experiment conditions. The Debye length tends to exceed the mean free path for a sufficiently low neutral pressure (controls $l$) or high plasma density (controls $\nuLB$ and $\eta$). When $\DebL/\lambda_\mrm{e{\text -}e}^\mrm{LB+IE}>1$, we can consider electrons are collisional enough to the extent that the instability causes a shifted Maxwellian EVDF from a certain point in the presheath, and even inside the sheath region that scales with $\sim\DebL$, as $\nuIE$ grows exponentially toward the wall. The yellow asterisks represent the upper limit of the achievable plasma density in our discharge system for corresponding neutral gas pressure.  We chose to scan the experimental conditions for $p_\mrm{Ar}=0.1$~mTorr and varying plasma density which is roughly indicated by a magenta arrow in the figure. This is because the condition $p_\mrm{Ar}=0.1$~mTorr gives wide range of value on $\DebL/\lambda_\mrm{e{\text -}e}^\mrm{LB+IE}$ for both less and more than unity, which will provide a clear physical differences between them if there are any.

For a plasma condition where strong IE collisions are expected by flowing ions toward immersed biased wall, which is the case for $\phib\lesssim(\phip-\Te)$, we can expect to have the shifted Maxwellian EVDF. Based on the theoretical background using a fluid approach, the amount of collected electrons will be different from the conventional current. The factor $\Te$ in the inequality arises from the fact that the ion fluid speed at the sheath edge $u_\mrm{i,se}$ approaches zero when the bias is close to the plasma potential\cite{Baalrud_2011_RN651}. For $(\phip-\Te)\lesssim\phib\lesssim\phip$, the EVDF is expected have an intermediate shape between the truncated (or cone-shaped in 3D velocity space) and the flowing Maxwellian. For $\phip\le\phib$, the wall collects the random flux of electrons for half-Maxwellian EVDF giving the conventional electron saturation current. In this work, we do not consider streaming instabilities between ions and electrons that could drive electrons more collisional, as the rate of collisions are not evident to be large enough as required.

The electron current $\Ie$ towards the wall for strongly collisional electrons is then predicted as
\begin{equation}
\Ie= \left\{
\begin{array}{cl}
	\Iesmax \exp \left[-\frac{\left(\phi_{p}-\phi_{b}\right)}{\Te} \right] \hfill
		\text{for} \; \phib\lesssim\phip-\Te, \\
	\Iestran \exp \left[-\frac{\left(\phi_{p}-\phi_{b}\right)}{\Te} \right] \hfill
		 \text{for} \;\phip-\Te\lesssim\phib\lesssim\phip, \\
	\Iestrunc \hfill \text{for} \;\phib\ge\phip.
\end{array}\right.
\label{eq:ee_Ie_IE}
\end{equation}
Here, $\Iesat$ represents electron saturation current $\Iesat=e\Ap\Gamma_\mrm{e}$, where $e$ is the electron charge, $\Ap$ is the surface area of the wall, and $\Gamma_\mrm{e}$ from equation (\ref{eq:ee_Ie}). The subscripts on $\Iesat$ in equation (\ref{eq:ee_Ie_IE}) indicate the different shapes of the EVDF, which are such that `M' denotes a case for the shifted Maxwellian model, `T' the truncated model, and `tr' the transitional state between the two. Each subscripts corresponds to a different value for $\kape$, where we expect
\begin{equation}
	\kapetrunc(=0.4)\le\kappa_\mrm{e,tr}\le\kapemax(=0.55).
	\label{eq:ee_kape}
\end{equation}
It should be noted that instead of a value 0.6\cite{Guittienne_2018_RN137}, we use $\kapemax=0.55$ as this value represents the actual density factor at the sheath edge with respect to the bulk as that of ions, which is originated from the presence of an ion presheath and a finite size of the wall\cite{Sheridan_2000_RN706}.

\begin{figure}\centering
\includegraphics[width=\linewidth]{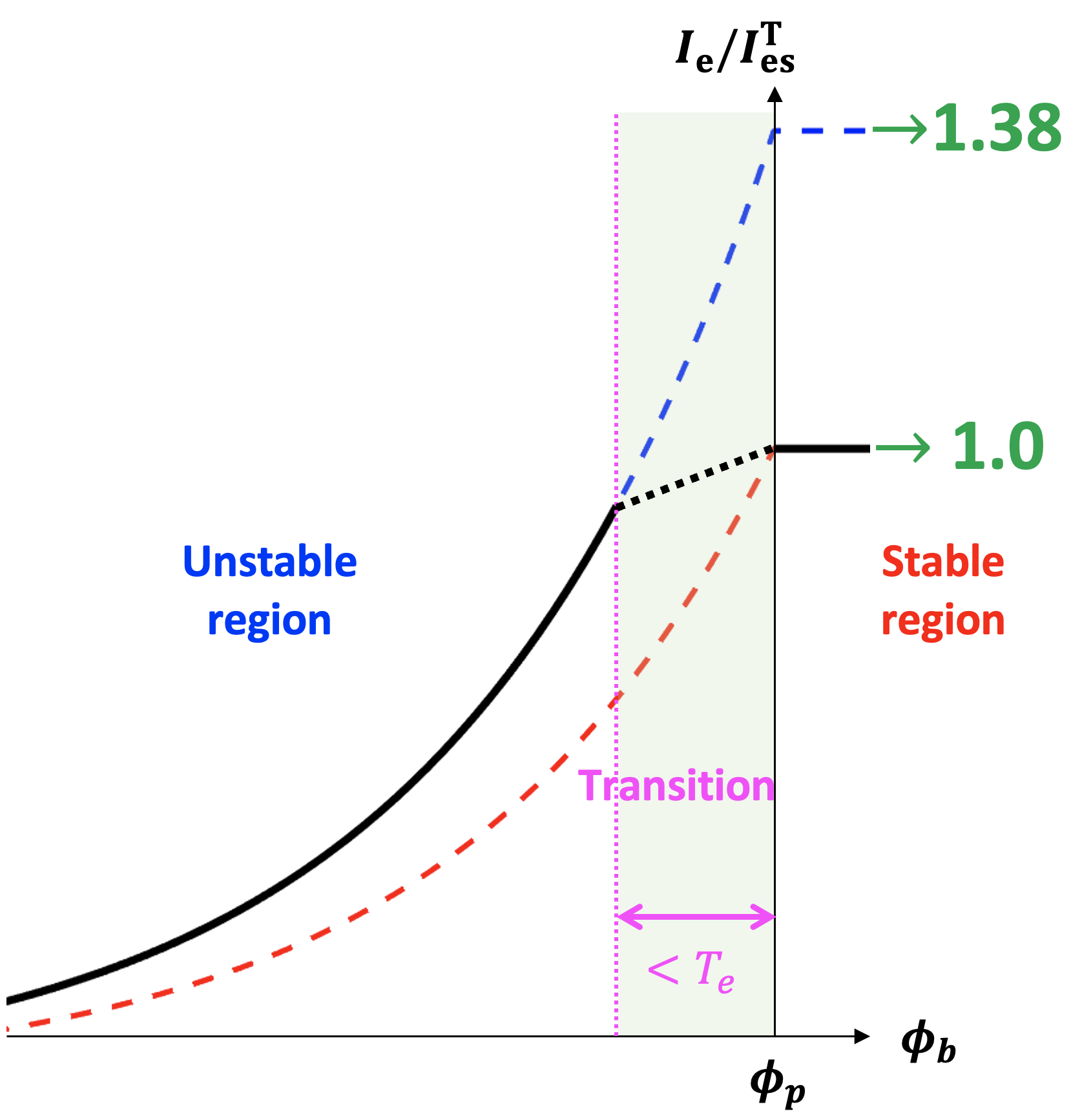}
\caption{A predicted \textit{I--V} characteristic measured by an ideal edge effect eliminated Langmuir probe with current normalized by $\Iestrunc$, in the presence of ion-acoustic instability with strongly enhanced e--e collisions in front of the probe wall.}
\label{fig:ee_intro}
\end{figure}

The shape of the \textit{I--V} characteristic described in equations (\ref{eq:ee_Ie_IE}) and (\ref{eq:ee_kape}) is illustrated in figure \ref{fig:ee_intro} with black solid lines, where the abscissa is the probe bias $\phib$ and the ordinate is the electron current normalized by $\Iestrunc$. As described earlier, the domain of the wall bias $\phib$ is divided into three regions when the plasma is in a condition that can establish strong IE collisions. They are an unstable region ($\phib\lesssim\phip-\Te$) where the IE collisions lead to the factor $\kapemax$, a stable region ($\phip\le\phib$) with collisionless electrons having $\kapetrunc$, and a transition region ($\phip-\Te\lesssim\phib\lesssim\phip$) between the two with $\kappa_\mrm{e,tr}$. The blue and red dashed lines indicate curves corresponding to constant $\kapemax$ and $\kapetrunc$, respectively. The extrapolated value of the curve from the unstable region at $\phip$ coincides with $(\Iesmax/\Iestrunc)=(\kapemax/\kapetrunc)=(0.55/0.4)\approx1.38$.

\begin{figure}\centering
\includegraphics[width=\linewidth]{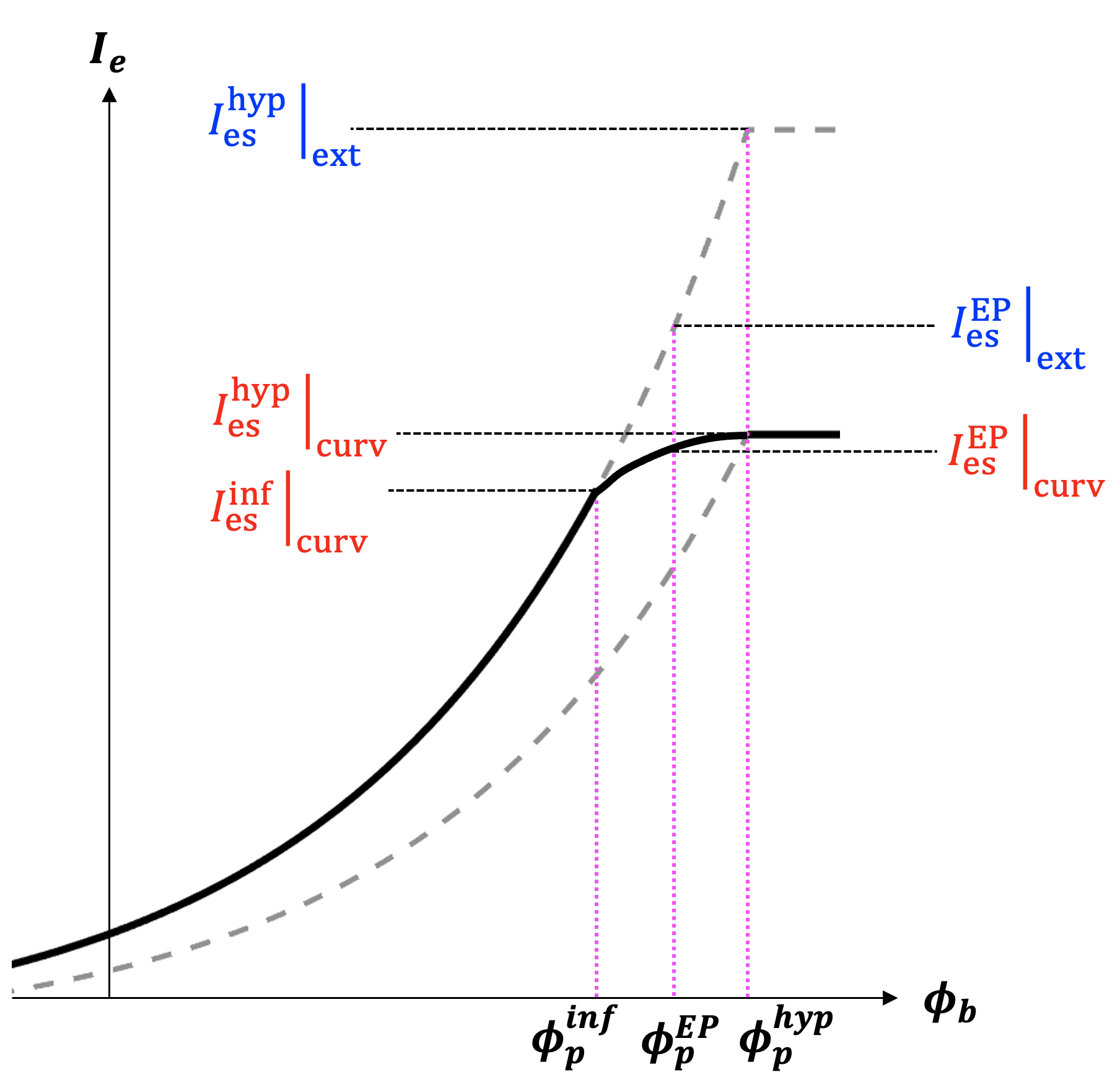}
\caption{An expected \textit{I--V} characteristic that will be measured in this work and defined notations. $\phip^\mrm{(est)}$ denotes estimated critical potentials for different methods, and as an example, an inflection point (inf), a plasma potential measured by emissive probe (EP), and a hypothetical potential (hyp) is shown. Corresponding measured probe current is represented by $\left. I_\mrm{es}^\mrm{(est)} \right|_\mrm{curv}$, and current extrapolated from the domain of the region $\phib \ll \phip$ by $\left. I_\mrm{es}^\mrm{(est)} \right|_\mrm{ext}$.}
\label{fig:ee_IV}
\end{figure}

Figure \ref{fig:ee_IV} illustrates an expected \textit{I--V} characteristic that will be actually measured in this work, shown with notations on currents and defined potentials. The estimated critical potentials are denoted by $\phip^\mrm{(est)}$. To give an example, an inflection point (inf), a potential measured by emissive probe (EP), and a hypothetical potential (hyp) are shown in the figure with corresponding electron currents. The corresponding estimated electron saturation current $\Iesat^\mrm{(est)}$ on the measured \textit{I--V} characteristic is expressed as $\left.\Iesat^\mrm{(est)}\right|_\mrm{curv}$. We also evaluate the electron current values on the extrapolated curve from the exponential part ($\phib\ll\phip$) for the given potentials $\phip^\mrm{(est)}$, and these values are expressed as $\left.\Iesat^\mrm{(est)}\right|_\mrm{ext}$.

For the ideal \textit{I--V} characteristic such that an emissive probe gives the true plasma potential, it is predicted as follows: (1) For the case of a plasma with collisionless electrons over the whole spatial domain, $\phipinf=\phipep$ and $\IesEPext/\IesEPcurv=1$. (2) When the electrons are sufficiently collisional leading to the shifted Maxwellian EVDF at the presheath and sheath region, $\phiphyp=\phipep$ such that $\IesEPext/\IesEPcurv=\kapemax/\kapetrunc=1.38$. (3) For the intermediate condition between (1) and (2), we predict that $\phipinf<\phipep<\phiphyp$ giving $1<\IesEPext/\IesEPcurv<1.38$.

Since the estimation of the plasma potential even for the collisionless case can give slight deviation from the true potential in real experiments, we developed several methods on estimating critical potential points to navigate their evolution when changing the plasma conditions where the methods for their estimation will be described in section \ref{sec:ee_Vp}. There also can exists undesired factors that may induce additional roundness of the knee or unsaturated currents that may disturb our analyses. Thus, we utilize cutoff probe measurements to evaluate $\kape^\mrm{(est)}$, where $\kape^\mrm{(est)}$ is a factor for the curve in the region $\phib\lesssim(\phip-\Te)$. Detailed description on estimating $\kape^\mrm{(est)}$ will be elaborated in section \ref{sec:ee_results}.
 
\section{Experimental setup and procedure}\label{sec:ee_setup}
The experiments are conducted in a cylindrical multidipole chamber of 1~m in length and 0.6~m in diameter\cite{Lim_2020_RN239, Lim_2024_RN961}. The vacuum chamber has a base vacuum pressure of approximately $2\times10^{-6}$~Torr. The chamber is filled with Ar gas in fixed working pressure of 0.1~mTorr, and the plasmas are generated by energetic primary electrons emitted from hot thoriated tungsten filaments biased to $-80$~V.

The EERP is a flat tantalum disc having a diameter of 10~mm, and is surrounded by a grounded guard-ring probe of 30~mm in outer diameter. The gap between the probes is approximately $0.05$~mm. The total surface area of the center and guard-ring probes is small enough not to induce non-ambipolar diffusion that affects the global plasma properties\cite{Amemiya_1991_RN987, Baalrud_2007_RN478}. The probes are placed on the center of a Al$_2$O$_3$ disc insulator. The probes are partly apart from the alumina disc by plain washers with 0.8~mm in thickness to prevent possible electrical connection between the two probes by contamination on the insulator surface.

\begin{figure}\centering
\includegraphics[width=\linewidth]{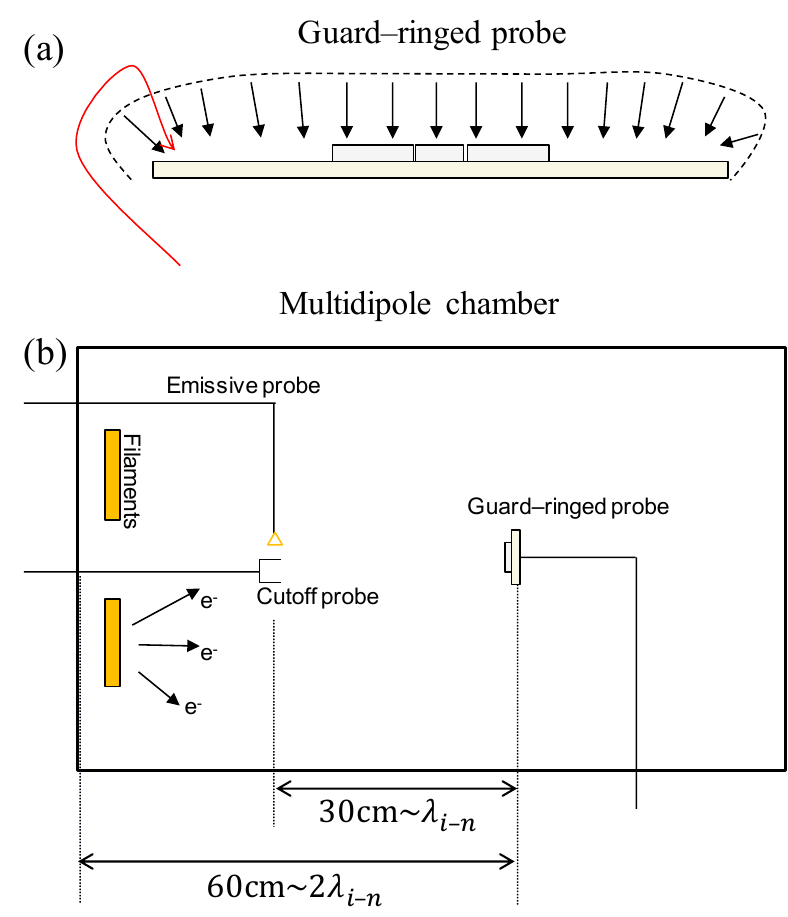}
\caption{(a) The schematic drawing of the guard-ringed probe EERP where a dashed line represents sheath edge and black arrows the direction of electron flux. (b) The illustration of overall experimental setup.}
\label{fig:ee_setup}
\end{figure}

Figure \ref{fig:ee_setup}(a) illustrates a schematic drawing of a structure holding EERP with the sheath edge indicated by the dashed line and the expected electron flow entering the sheath edge by black arrows. The red arrow represents the potential orbit of ions\cite{Stamate:2005kj}, which could distort measurements if the diameter of the insulator disc is not large enough. It should be noted that the sheath edge will not be perfectly flat-shaped since we only sweep the voltage to the center probe while leaving the surrounding guard-ring probe grounded. It has been confirmed that this configuration effectively reduces edge effects and achieves a similar rate of increasing unsaturated electron current for $\phib>\phip$. Biasing both probes simultaneously would rather have possibility on perturbing the entire plasma and distorting the \textit{I--V} characteristics.

While \textit{I--V} characteristics are obtained by the EERP, we employ emissive and cutoff probes to measure absolute plasma potentials and densities, respectively. These measurements provide reference plasma parameters for simultaneously measured data from the EERP. As depicted in figure \ref{fig:ee_setup}(b), the emissive and cutoff probes are located at the middle of the filaments and the EERP, with a distance of $\sim30$~cm from them in opposite directions. This arrangement allows for measurements to be taken in the bulk plasma region.

\begin{figure}\centering
\includegraphics[width=0.9\linewidth]{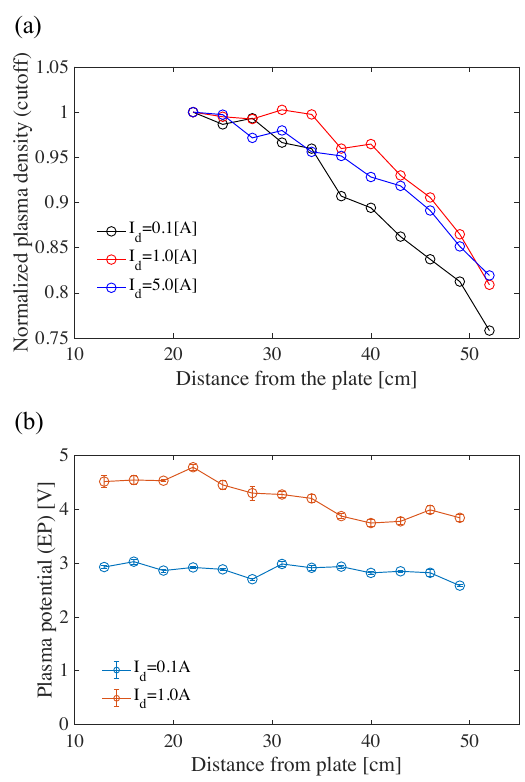}
\caption{The profiles of the plasma properties generated in the cylindrical multidipole chamber without the EERP structure in axial direction, where 0~cm is the position of the EERP when it is installed.  (a) Normalized plasma densities measured by a cutoff probe, and (b) plasma potentials measured by an emissive probe. The result shows that the cutoff probe and the emissive probe measurements can be used as a reference values for measurements from the EERP.}
\label{fig:ee_spatial}
\end{figure}

To ensure the validity of our experiments that will correlate the data measured from EERP with measurements at the bulk region by emissive and cutoff probes, we measured the axial distributions of the plasma density and potential without the EERP structure. This is shown in figure \ref{fig:ee_spatial} for different discharge currents, which is the key control parameter in this work. The shape of the density distribution remains nearly unchanged with varying discharge current, indicating that the measured density obtained by the cutoff probe ($\np^\mrm{co}$) can be used as a reference density by multiplying it with a constant heuristic factor $f_\mrm{cal}$, i.e., $\np=f_\mrm{cal}\np^\mrm{co}$ where $\np^\mrm{co}$ is the density measured by the cutoff probe. The distributions of the plasma potential measured by the emissive probe ($\phipep$) are found to be nearly uniform under our experimental circumstances. Note that the plasma in our device is typically uniform along the radial direction\cite{Lim_2020_RN239}.

We collected \textit{I--V} characteristics from EERP together with measurements from the emissive and cutoff probes, varying the discharge current from 0.02~A to 7~A. This procedure is to explore plasmas over the range that indicated as a magenta arrow in figure \ref{fig:IE}. If our predictions based on the theories are valid, the feature of the electron current will be observed, i.e., a gradual evolution from the conventional model to the new predicted model represented by equation (\ref{eq:ee_Ie_IE}) and the behavior of $\kape$. It should be noted that the emissive and cutoff probes are distanced 30~cm away from the EERP, which is comparable to $\lambda_\mrm{i{\text -}n}$, the mean free path of ion--neutral collisions. This fact indicates that the emissive and cutoff probes are inserted in the bulk region, and they do not hinder the growth of ion-acoustic instabilities towards the EERP, at least within a distance of $\lambda_\mrm{i{\text -}n}$.

\section{Estimation of critical potentials in \textit{I--V} characteristic}\label{sec:ee_Vp}

The predictions we have made in section \ref{sec:ee_expect} indicate that if the electron--electron collisions are sufficiently enhanced by ion-acoustic instability, the value $\kape$ reduces in the transition region $(\phip-\Te)\lesssim\phib\lesssim\phip$ as $\phib$ increases towards $\phip$, rounding off the the knee of the \textit{I--V} characteristics. This rounded knee causes an underestimation of the plasma potential for approximately $\Te$ at maximum when using one of the most common method\cite{Hershkowitz:1989}, the inflection point method. The discrepancy between the true plasma potential and the inflection point of the curve, which is believed to be provided by the emissive probe measurements, is the important evidence for verifying our predictions. Note that we used the inflection point method in the limit of zero emission for the emissive probe diagnostics\cite{Sheehan_2011_RN741}.

The knee of the \textit{I--V} characteristic can be rounded off due to several reasons, even when the electrons are collisionless\cite{Hershkowitz:1989, Tichy_1997_RN897}. One major probable cause is surface contamination, so we cleaned the EERP by heating the probe before conducting the experiments. Nonetheless, we presume that the amount of underestimation of the plasma potential for the inflection point will still present by the inherent roundness, which cannot be completely eliminated, but remains constant over our experiment conditions. Typically, this underestimation is negligible for plasmas generated at low neutral pressure\cite{Li_2020_RN439} compared to the maximum width of the predicted transition region ($\sim\Te$).

\begin{figure}\centering
\includegraphics[width=\linewidth]{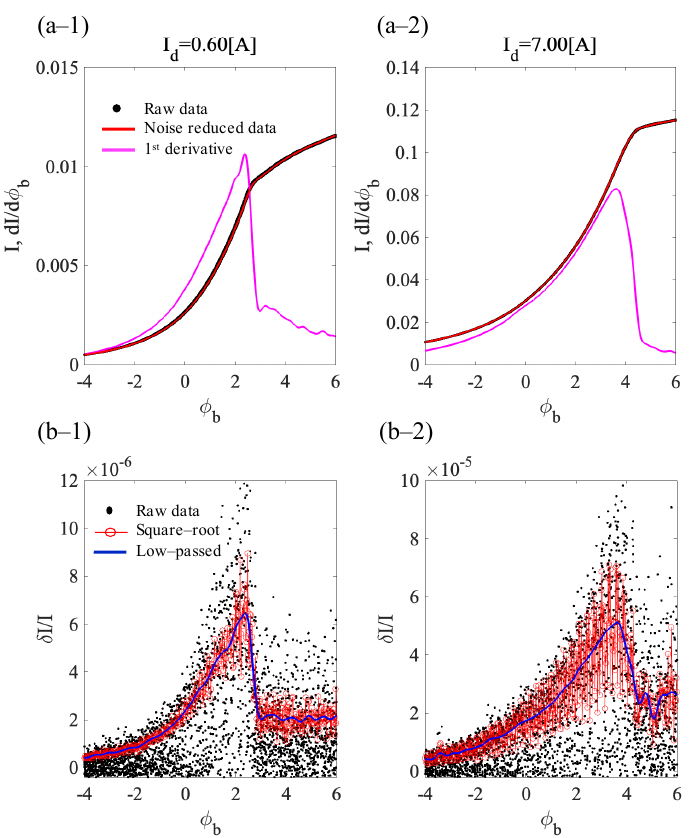}
\caption{Examples of raw and their processed \textit{I--V} characteristics measured by the guard--ringed probe at discharge current 0.60 A (a$\vert$b--1) and 7.0 A (a$\vert$b--2). (a) Raw data (black dots), and its noise--reduced data (red) and the 1st derivative (magenta) is shown. In (b), the current fluctuations evaluated by difference between raw and noise--reduced data in (a) is shown as black dots, with its root mean square (red circles) and low--passed (blue) data.}
\label{fig:ee_fluct}
\end{figure}

Because the inflection point method is not the ideal method, we attempted to estimate critical points around the plasma potential using different approaches. One of these methods is the `maximum curvature' method. Just above the inflection point, the first derivative of the curve sharply decreases around the rounded knee. The point of steepest descent in the first derivative corresponds to the maximum curvature of the \textit{I--V} characteristic and is referred to as $\phipmxc$. The second method is the `maximum fluctuation' method. Figure \ref{fig:ee_fluct} shows examples of the raw \textit{I--V} characteristics (black dots) and their derivative (magenta) in (a), and their current fluctuations in (b) for two discharge currents, 0.60 and 7.0 A. The fluctuations are assessed by calculating the difference between the raw data and noise-reduced, ensemble averaged data. These fluctuations are filtered and subjected to a square root to determine the point of maximum fluctuation, denoted by $\phipmxf$.

We also estimated the point where the true potential is anticipated when electrons are in a shifted Maxwellian EVDF. This point corresponds to when $\Iesmax=1.38\Iestrunc$, denoted by $\phiphyp$, as depicted in figure \ref{fig:ee_IV}. According to the arguments we presented in section \ref{sec:ee_expect}, when the electrons are collisionless, $\phipep$ will be close to $\phipinf$, and as the plasma density increases during the experimental process described in section \ref{sec:ee_setup}, $\phipep$ will approach $\phiphyp$ as the electrons become collisional. To be more specific, our predictions indicate that $\phipep\sim\phiphyp$ when $\DebL/\lambda_\mrm{e{\text-e}}^\mrm{LB+IE}\gtrsim1$.

\begin{figure}\centering
\includegraphics[width=\linewidth]{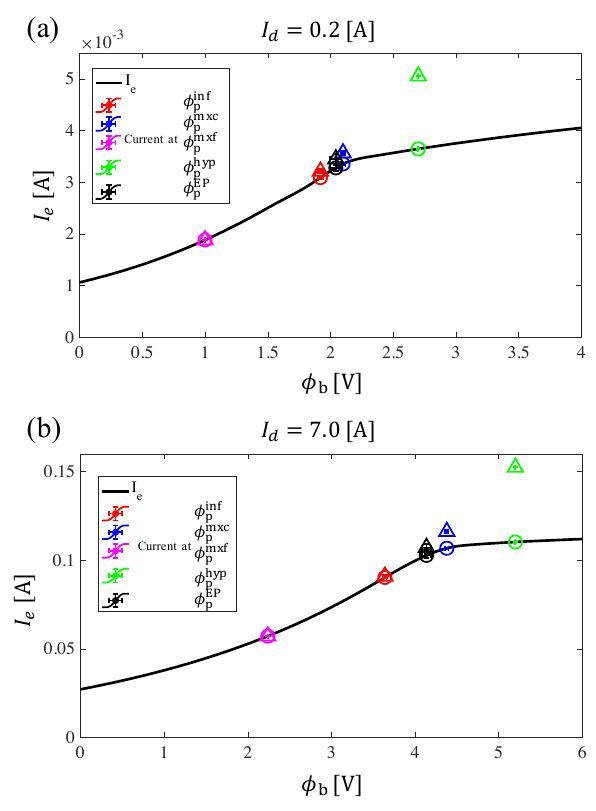}
\caption{Examples of electron currents $\Ie$ as a black solid lines and corresponding electron saturation currents for plasma potentials estimated by different methods. Circles denotes the point on the curve, $\left. \Iesat^\mrm{(est)} \right|_\mrm{curv}$, and triangles the point extrapolated from the electron current at low probe bias ($\phib \lesssim \phip-\Te$), $\left. \Iesat^\mrm{(est)} \right|_\mrm{ext}$.}
\label{fig:ee_Vp}
\end{figure}

Figure \ref{fig:ee_Vp} shows examples of the estimated plasma potentials along with their corresponding $\left.\Iesat^\mrm{(est)}\right|_\mrm{curv}$ (circles) and $\left.\Iesat^\mrm{(est)}\right|_\mrm{ext}$ (triangles) for the discharge current $I_\mrm{d}=0.2$ and $7.0$~A in (a) and (b), respectively. In the figure, black solid lines represent the ensemble-averaged and low-passed \textit{I--V} characteristics. Note that the electron current is obtained by subtracting the linearly fitted ion current from the \textit{I--V} characteristic where its magnitude is very small compared to the electron current because of the edge-effect reduction by the EERP structure.

\section{Results and discussions}\label{sec:ee_results}

As described in section \ref{sec:ee_setup}, we varied the discharge current from 0.02 to 7.0~A which corresponds to an approximate range of plasma density from $1.7\times10^8$ cm$^{-3}$ to $2.5\times10^{10}$ cm$^{-3}$. The plasma density is determined by using the electron saturation current based on the conventional probe theory, i.e., from $\Iesinfcurv$, which may slightly overestimate the density if our prediction for the shifted Maxwellian EVDF applies. However, this does not impact the conclusion of this work. Throughout the experiments, the effective electron temperature $\Teff$ increased from 0.8~eV to 2.8~eV, where the electrons are bi-Maxwellian distributed and $\Teff$ is evaluated by taking the harmonic mean of the two-temperature electron fractions. It should be noted that the estimated plasma density and electron temperature are employed to determine $\DebL$ and $\lambda_\mrm{e{\text -}e}^\mrm{LB+IE}$ in this section.

\begin{figure}
\includegraphics[width=\linewidth]{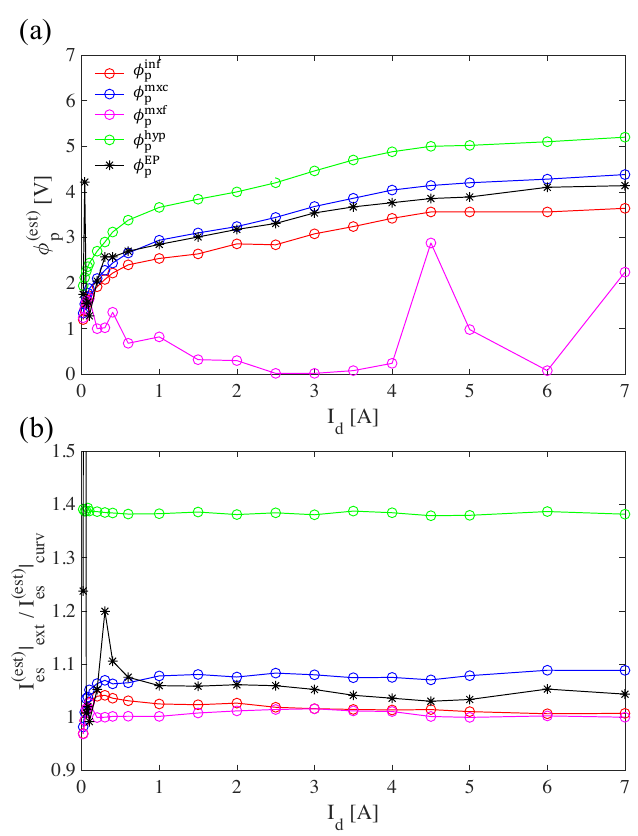}
\caption{The evolution of (a) plasma potentials estimated by different methods and (b) corresponding extrapolated electron saturation currents normalized by $\left.\Iesat^\mrm{(est)}\right|_\mrm{curv}$ for increasing discharge current $I_\mrm{d}$.}
\label{fig:ee_VpIes}
\end{figure}

Figure \ref{fig:ee_VpIes} shows the evolution of the estimated critical potentials in (a) and extrapolated electron saturation currents $\left.\Iesat^\mrm{(est)}\right|_\mrm{ext}$ normalized by $\left.\Iesat^\mrm{(est)}\right|_\mrm{curv}$ in (b). It is clearly seen that the estimated plasma potentials follow the order $\phipmxf<\phipinf<\phipep<\phipmxc<\phiphyp$, except for the low density plasmas with discharge conditions $I_\mrm{d}\lesssim0.5$~A. Similarly, the corresponding normalized extrapolated electron saturation currents in (b) exhibit the same order and remain nearly constant over different discharge conditions. The arbitrariness observed in the results for the low density plasmas is presumably attributed to small signal-to-noise ratio of our data. Nevertheless, the consistent sequence of the potentials and normalized saturation currents observed in our experiments validates the accuracy of our measurement.

\begin{figure}\centering
\includegraphics[width=\linewidth]{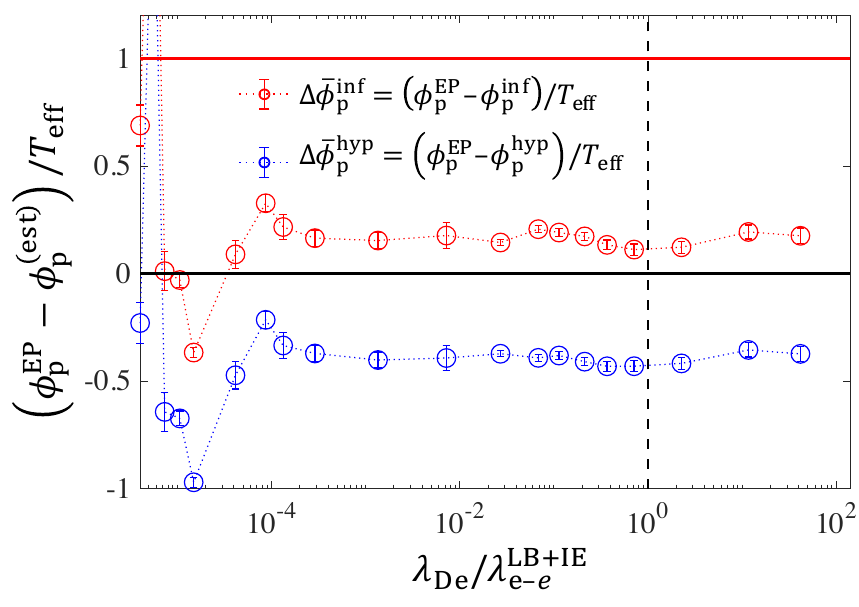}
\caption{The difference between plasma potentials refer to the potentials estimated by the emissive probe, and inflection point method (red circles) and hypothetical potential (blue circles) as a function of $\DebL/\lambda_\mrm{e{\text -}e}^\mrm{LB+IE}$. The red solid denotes where we expect $(\phipep-\phipinf)$ be reached for sufficient e--e collisions which is approximately coincides with the right side of the vertical dashed line.}
\label{fig:ee_comparison}
\end{figure}

Based on the observed ordered feature of the estimated critical potentials for different methods, we concluded that it is reasonable to compare the inflection point $\phipinf$ with true plasma potential $\phipep$ along with $\phiphyp$ to proceed our analysis to capture the evolution of parameters. Figure \ref{fig:ee_comparison} presents the differences of the estimated potential differences shown in figure \ref{fig:ee_VpIes}(a) as a function of $\DebL/\lambda_\mrm{e{\text -}e}^\mrm{LB+IE}$ to clearly demonstrate the electron collisionality. The behavior for $\DebL/\lambda_\mrm{e{\text -}e}^\mrm{LB+IE}\lesssim10^{-4}$ corresponds to the experimental noise for $I_\mrm{d}\lesssim0.5$~A. It is noteworthy that both potential differences, $\Delta\bar{\phi}_\mrm{p}^\mrm{inf}\equiv(\phipep-\phipinf)/\Teff$ and $\Delta\bar{\phi}_\mrm{p}^\mrm{hyp}\equiv(\phipep-\phiphyp)/\Teff$, remain constant over the range of $10^{-4}<\DebL/\lambda_\mrm{e{\text -}e}^\mrm{LB+IE}<10^2$. This result indicates that even when electron--electron collisions are significantly enhanced and their mean free path is much smaller than the Debye length, they do not exhibit the predicted transition region in the \textit{I--V} characteristic represented by the width of the rounded knee. These findings conflict with our predictions which anticipated $0<\Delta\bar{\phi}_\mrm{p}^\mrm{inf}\ll1$ for $\DebL/\lambda_\mrm{e{\text -}e}^\mrm{LB+IE}\ll1$, and as the electron mean free path decreases, gradual increment of $\Delta\bar{\phi}_\mrm{p}^\mrm{inf}$ with saturation between $0.5$ and $1$ above a certain threshold over $\DebL/\lambda_\mrm{e{\text -}e}^\mrm{LB+IE}\sim1$. Similarly, we expected $\Delta\bar{\phi}_\mrm{p}^\mrm{hyp}<0$, and likewise eventually saturate around $\Delta\bar{\phi}_\mrm{p}^\mrm{hyp}\sim0$ above the collisionality threshold.

We also estimated $\kape$ by using reference density from the cutoff probe and the values of the extrapolated electron saturation currents $\left.\Iesat^\mrm{(est)}\right|_\mrm{ext}$, the method types denoted in the superscript. To obtain $\kape$, the saturation currents were separated into the hot and cold electron fractions and divided by $e\Ap\vthe$ yielding values that imply $\kape^\mrm{(est)} \np$. These values were than divided by $1.6\np^\mrm{co}$, which is the density measured by the cutoff probe $\np^\mrm{co}$ multiplied by the calibration factor $f_\mrm{cal}$=1.6. Here, we determined the value of $f_\mrm{cal}$ based on the resultant $\kape^\mrm{inf}$ to coincide with $\kapetrunc$, which aligns with our predictions. It should be noted that $\kape^\mrm{EP}$ represents a true $\kape$ for the region $\phib\lesssim(\phip-\Te)$ as $\phipep$ is believed to be the correct plasma potential. Depending on the experiment conditions, the region $\phib\lesssim(\phip-\Te)$ can be either stable or unstable in regard to the ion-acoustic instability, thus the factor $\kape^\mrm{EP}$ governing this region will have a value within $\kapetrunc\le\kappa_\mrm{e,tr}\le\kape^\mrm{EP}\le\kapemax$.

\begin{figure}\centering
\includegraphics[width=\linewidth]{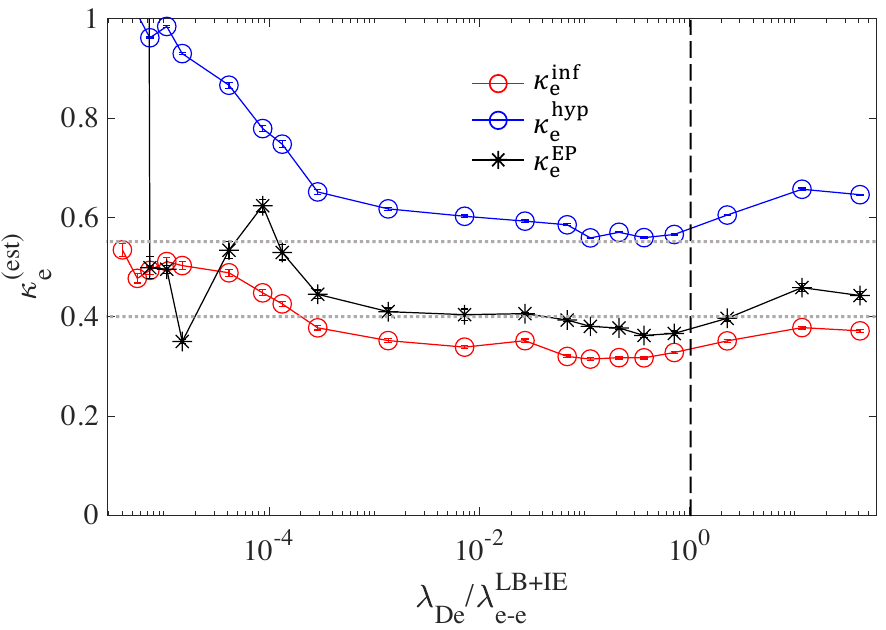}
\caption{The evaluated $\kape^\mrm{(est)}$ using the extrapolated electron saturation currents for different methods on estimating plasma potentials as a function of $\DebL/\lambda_\mrm{e{\text -}e}^\mrm{LB+IE}$. Two dotted horizontal lines correspond to $\kapetrunc$ and $\kapemax$, and a vertical dashed line at  $\DebL/\lambda_\mrm{e{\text -}e}^\mrm{LB+IE}=1$ denotes where $\kape^\mrm{EP}$ is expected to exhibit certain transition.}
\label{fig:ee_kape}
\end{figure}

The results on the estimation of $\kape^\mrm{(est)}$ shown in figure \ref{fig:ee_kape} further demonstrate a discrepancy between the predictions and observations. Again, points for $\DebL/\lambda_\mrm{e{\text -}e}^\mrm{LB+IE}\lesssim10^{-4}$ are dummy data because of the experimental error. According to the predictions associated with equation (\ref{eq:ee_kape}), we expected the estimated $\kape^\mrm{EP}$ will be comparable to $\kape^\mrm{inf}\approx\kapetrunc$ for $\DebL/\lambda_\mrm{e{\text -}e}^\mrm{LB+IE}\ll1$, and increase as electrons become collisional. Again, above a certain threshold around $\DebL/\lambda_\mrm{e{\text -}e}^\mrm{LB+IE}\sim1$, we anticipated $\kape^\mrm{EP}$ would approach $\kape^\mrm{hyp}\approx\kapemax$. However, we did not observed outcomes that aligned with our predictions, likewise in the manner that we examined the results shown in figure \ref{fig:ee_Vp}.

One might raise a suspicion on that the intensity of the ion-acoustic instability was not sufficiently established. In this work, we used a length of $l=2\lambda_\mrm{i{\text -}n}$ from the instability onset to the EERP to determine $\nuIE$ in designing our experiment, as described in section \ref{sec:ee_expect} and figure \ref{fig:IE}. This assumption may be insufficient since the axial displacement between the EERP and the chamber wall is approximately $2\lambda_\mrm{i{\text -}n}$ (see figure \ref{fig:ee_setup}(b)), and the acceleration of the ion flow towards the EERP could potentially initiated from the middle region. However, even when we considered a distance of $l=\lambda_\mrm{i{\text -}n}$, we still have $\DebL/\lambda_\mrm{e{\text -}e}^\mrm{LB+IE}\approx12$ for the most dense condition in our experiments. This value is significantly large enough to justify the assumption of the shifted Maxwellian electrons within substantial space domain as the Debye length corresponding to the condition is approximately $0.08$~mm.

Since the IE collisions are present at the presheath and sheath region and its collisionality is exponentially growing toward the wall, it is true that the IE collisions do not make electrons highly collisional over the whole spatial domain of plasmas. However, even when the spatial regions are divided into electrons having truncated and flowing Maxwellian, as far as electrons reach the wall with its thermal velocity as the fluid model predicts\cite{Guittienne_2018_RN137}, our conjecture is still valid in terms of the fact that $\kape$ varies.

In consideration on the concrete theoretical background we have been concerned with, our results are quite surprising. The presence of enhanced collisions for ion--ion collisions due to the ion-acoustic instability has been confirmed in many experimental studies\cite{Baalrud_2015_RN637, Yip_2015_RN164, Severn_2017_RN63}. The enhancement principle of collisions between electrons stems from the same theoretical background, i.e., the collective nature of plasmas. Therefore, IE collisions for electrons are highly probable to present in our experiments. The fluid approaches applied to the electrons, along with the theory of the electron Bohm criterion, also appear to be mathematically sound.

Our results show that the electron current collected by the wall is explained by the conventional wisdom represented by the truncated EVDF model and the random flux of electrons when $\phib=\phip$, for both collisionless and collisional cases. The next step in confirming the predictions presented in this work is to directly measure the EVDF in the sheath and presheath using Thomson scattering diagnostics across the range of discharge conditions considered in this work to figure out the presence of flowing Maxwellian EVDF. This is a quite challenging task, as achieving conditions of high rate of electron--electron collisions typically requires a dense plasma, resulting in a thin sheath. The particle-in-cell (PIC) simulation also can be a feasible approach to resolve the discrepancies, since the PIC method can capture the ion-acoustic IE collisions\cite{Scheiner_2019_RN373}, at least for ions. Difficult point is that the size of the cell must be less then $0.5\DebL$ where our experiment conditions have $\DebL\sim10^{-4}$~m in minimum, while the simulation domain must cover the characteristic length of presheath. This corresponds to $\sim10^8$ of cell grid when we employ equidistant 2D Cartesian grid, and it requires high computational power. We believe that verifying the shape of the EVDF near the sheath will open a new chapter in the experimental study related with recent theoretical works on highly collisional presheath and sheath, and also on electron sheath.

\section{Conclusion}\label{sec:ee_conclusion}

We established predictions on electron currents collected by the wall for the ion sheath where the wall bias is equal or less than the plasma potential. The predictions are based on different outcomes derived by kinetic and fluid models. Instead of the factor $\kape=1/\sqrt{2\pi}$ that is multiplied to the electron current for the conventional kinetic model, fluid model predicts that the factor is $\kape=\mrm{e}^{-1/2}$, where the fluid approach can be applied if the electrons are highly collisional resulting in a shifted Maxwellian velocity distribution function. The experimental design exploring transitions between the two distinguishing consequences is developed in consideration of the theory on ion-acoustic instability-enhanced electron-electron collisions. We utilized an edge-effect reduced Langmuir probe to remove the sheath expansion and edge effects from the \textit{I--V} characteristic. The prediction for collisional electrons is represented by an extended width of a rounded knee of the curve compared to the conventional collisionless model, and unique methods to capture critical points around the knee is suggested. To improve validity of our experiment, we employed emissive and cutoff probe diagnostics to provide reference parameters in the bulk plasmas for the edge-effect reduced Langmuir probe data. Our observation indicates that the electron current collected by the probe is explained by the conventional wisdom even for highly collisional circumstances. We suggest experiments employing a Thomson scattering diagnostics and particle-in-cell simulations that can be done for making further advances regarding our work.


\section*{Acknowledgement}

\section*{References}
\bibliographystyle{unsrt}
\nocite{}
\bibliography{eeIe_v0.bib}

\end{document}

%% file: my_macro.tex
\usepackage{color}

\newcommand{\trm}{\textrm}
\newcommand{\mrm}{\mathrm}

